%% file: skeleton.tex
\documentclass[a4paper]{PoS}

\input{aliases.tex}

\title{Measurement of angular correlations between \textit {D} mesons and charged particles in pp and p--Pb  collisions with ALICE at the LHC}

\ShortTitle{D meson-charged particle correlations}

\author{\speaker{Somnath Kar} 
	{on behalf of ALICE Collaboration}
	\\
        Variable Energy Cyclotron Centre, Kolkata, India\\
        E-mail: \email{somnath.kar@cern.ch}}

\abstract{We have studied the azimuthal correlations between $\Dzero$, $\Dplus$ and $\Dstar$ mesons and charged particles in pp collisions at $\sqrts=7~\tev$ and p--Pb collisions at $\sqrtsNN=5.02~\tev$ at the Large Hadron Collider. D mesons were reconstructed from their hadronic decays in the central rapidity region and in the transverse-momentum range 3 $\le p_{T} \le$ 16 $\gevc$, and they were correlated to charged particles reconstructed in the pseudo-rapidity range $|\eta| < 0.8$. A comparative study of the pp results with Monte Carlo Pythia studies and also with p--Pb results are presented here.}

\FullConference{7th International Conference on Physics and Astrophysics of Quark Gluon Plasma\\
		1-5 February , 2015\\
		Kolkata, India}
\begin{document}
\section{Introduction}
\label{sec:intro}
\input{introductionICPA.tex}
\section{D meson-charged particle azimuthal correlation analysis}
\label{sec:Analysis}
\input{DsignalExtraction.tex}
\section{Results}
\label{sec:Results}
\input{Results.tex}
\section{Summary}
\label{sec:conclusions}
\input{conclusions.tex}


\end{document}

%% file: aliases.tex
\newcommand{\sqrts}{\sqrt{s}}
\newcommand{\sqrtsNN}{\sqrt{s_{\scriptscriptstyle \rm NN}}}

\newcommand{\gevc}{\mathrm{GeV}/c}
\newcommand{\tev}{\mathrm{TeV}}

\newcommand{\pt}{p_{\rm T}}
\newcommand{\ptD}{p_{\rm T}({\rm D})}
\newcommand{\ptAssoc}{p_{\rm T}^{assoc}}

\newcommand{\DtoKpi}{{\rm D}^0 \to {\rm K}^-\pi^+}
\newcommand{\DtoKpipi}{{\rm D}^+\to {\rm K}^-\pi^+\pi^+}
\newcommand{\DstartoDpi}{{\rm D}^{*+} \to {\rm D}^0 \pi^+}

\newcommand{\Dzero}{{\rm D^0}}

\newcommand{\Dstar}{{\rm D^{*+}}}

\newcommand{\Dplus}{{\rm D^+}}

\newcommand{\vtwo}{v_{2}}

%% file: introductionICPA.tex
ALICE (A Large Ion Collider Experiment), one of the four main experiments at the LHC, is designed and optimised for the study of heavy-ion collisions. The main objective of the ALICE physics programme is to study the Quark Gluon Plasma (QGP), a state of matter in which quarks and gluons are not confined into hadrons, that is expected to be formed in high-energy collisions of heavy nuclei. Heavy quarks (charm and beauty), with large masses ($m_{c}$ $\approx$ 1.3 and $m_{b}$ $\approx$ 4.5 GeV/$c^{2}$), are well suited probes to study this state of matter as they are produced in the initial stages of the collision~\cite{cite1} via hard partonic scattering processes before the QGP is formed. So, they experience the full evolution of the medium and lose energy via both collisional and radiative processes~\cite{cite2, cite3, cite5} while interacting with the medium constituents, as supported by several experimental observations~\cite{cite4, STARHQEnloss, PHENIXHQEnloss}. Energy loss effects can be studied by comparing the yield of heavy-flavour particles in heavy-ion collisions to that in pp collisions. The study of heavy-quark production in pp collisions is also important to test perturbative QCD calculations. Measurements in proton-lead (p--Pb) collisions are also important to disentangle ``hot'' medium (QGP) effects in heavy-ion collisions from initial-state effects due to cold nuclear matter.\\
Two particle correlations are an efficient tool for investigating the particle production mechanism. Correlations involving heavy quarks in heavy-ion collisions can provide information to understand the heavy-quark energy loss. The $D-\bar{D}$ correlation studies were proposed at RHIC energies to understand the contributions of perturbative and non-perturbative QCD processes to the correlation functions~\cite{cite6, cite7}.\\
Azimuthal correlations between D mesons and charged particles in high-energy collisions can give a way to characterise charm production and fragmentation processes. 
The two dimensional correlations in azimuthal angle and pseudo-rapidity ($\Delta\varphi, \Delta\eta$) of D mesons and charged particles exhibit a "near-side" peak at $\Delta\varphi=0$ 
having D mesons as trigger particles, and an "away-side" peak at $\Delta\varphi$ = $\pi$. 
In heavy-ion collisions an additional structure emerges, which persists over large $\Delta\eta$ range on near and away side. The shape of the correlation in the azimuthal difference $\Delta\varphi$ can be described by expanding it in a Fourier series with coefficients $v_{n}$. The main component of the Fourier decomposition used to describe the anisotropy in the $\Delta\varphi$ distribution is the second order term, called elliptic flow or $v_{2}$. Recently we have seen long range ridge-like structures in high-multiplicity pp collisions at $\sqrts=7~\tev$~\cite{ppbdoubleridgeCMS} and in p--Pb collisions at $\sqrtsNN=5.02~\tev$~\cite{pPbdoubleridgeALICE, pPbdoubleridgeATLAS}. Similar results were obtained in central d--Au collisions at RHIC~\cite{dAuSTARlongrange}.\\
Data sample, analysis strategy including the D-meson signal extraction, the associated track selection criteria, and the corrections applied to measure the correlations between prompt D mesons and primary charged particles are reported in section 2. In section 3 results are discussed and a summary is given in section 4.

%% file: DsignalExtraction.tex
For the correlation study we have analysed $\approx$ 314$\times 10^{6}$ minimum bias pp events at $\sqrt s$ = 7 TeV and $\approx$ 100$\times 10^{6}$ minimum bias p--Pb events at $\sqrtsNN$ = 5.02 TeV. In case of pp, minimum-bias collisions are triggered by at least one hit in either V0A or V0C or in the Silicon Pixel Detector (SPD). For p--Pb collisions, the minimum bias trigger requires the signals from both the V0 detectors. 
\subsection{Analysis Strategy}
The azimuthal correlation between D mesons and charged particles is investigated using the following method. First the D ($\Dzero$, $\Dplus$, $\Dstar$ and their charge conjugates) mesons are reconstructed via their hadronic decay channels $\DtoKpi$(Branching Ratio of 3.88$\pm$0.05\%), $\DtoKpipi$(BR of 9.13$\pm$0.19\%), $\DstartoDpi$ (BR of 67$\pm$0.5\%)~\cite{pdg} and then they are correlated with the primary charged particles in the same event. For D-meson signal extraction we follow the same selection procedure as described here \cite{ppD2H7TeV, ALICEDmesonpPb}. The extraction of D-meson signal is based on the reconstruction of a secondary vertex, separated by few tens of micrometers from the primary vertex. The ALICE Inner Tracking System (ITS) provides the necessary resolution to identify the displaced secondary vertices. The Time Projection Chamber (TPC) and the Time Of Flight (TOF) detector allow identification of charged pions and kaons up to $p_{T}$ = 2 $\gevc$ which reduces the combinatorial background of uncorrelated pairs.\\
The raw yield of D mesons is extracted by fitting the invariant mass spectra with a Gaussian for the signal component and an exponential function for the background. Each D-meson candidate is correlated with all the primary charged particles, having $p_{T}$ > 0.3 $\gevc$ and pseudo-rapidity |$\eta$| < 0.8. In order to subtract the background due to the uncorrelated D candidates under the mass peak, the azimuthal correlation of background candidates in the sidebands of the D-meson peak with primary charged particles is used. The raw correlation ($\Delta\varphi, \Delta\eta$) between D mesons and charged particles is then corrected by the efficiency of trigger (D mesons) and associated particles (primary charged particles) using Monte Carlo simulated events.\\
In order to take into account the limited detector acceptance and spatial inhomogeneities, an event mixing approach is used. The corrected raw correlation from event mixing is nomalized by the number of triggers calculated via yield extraction using D-meson mass spectra. The per-trigger azimuthal correlation distribution is then obtained by multiplying with the purity of the primary particle sample. D mesons can originate from B-meson decays. In order to correct the raw correlation from correlated D mesons coming from B mesons we use a feed down correction method as described in~\cite{ppD2H7TeV}. Along with feed down correction various systematics uncertainties from D meson yield extraction, background subtraction, efficiency corrections, Monte Carlo closure test are calculated.\\
The fully corrected and normalised per-trigger D meson-charged particle azimuthal correlation is studied in different  D-meson $p_{T}$ ranges and with different $p_{T}$ intervals of associated primary charged particles.\\

%% file: Results.tex
The D meson-charged particle azimuthal correlation analysis was performed on three D-meson $\pt$ intervals in pp collisions (3 < $\pt$(D) < 5 $\gevc$, 5 < $\pt$(D) < 8 $\gevc$ and 8 < $\pt$(D) < 16 $\gevc$) and in two $\pt$ intervals in p--Pb collisions (5 < $\pt$(D) < 8 $\gevc$ and 8 < $\pt$(D) < 16 $\gevc$) and for different minimum associated track $\pt$ thresholds ($\ptAssoc$ > 0.3, 0.5 and 1 $\gevc$).\\
\begin{figure}[h!]
\centering
\includegraphics[width=14cm,height=0.33\textheight]{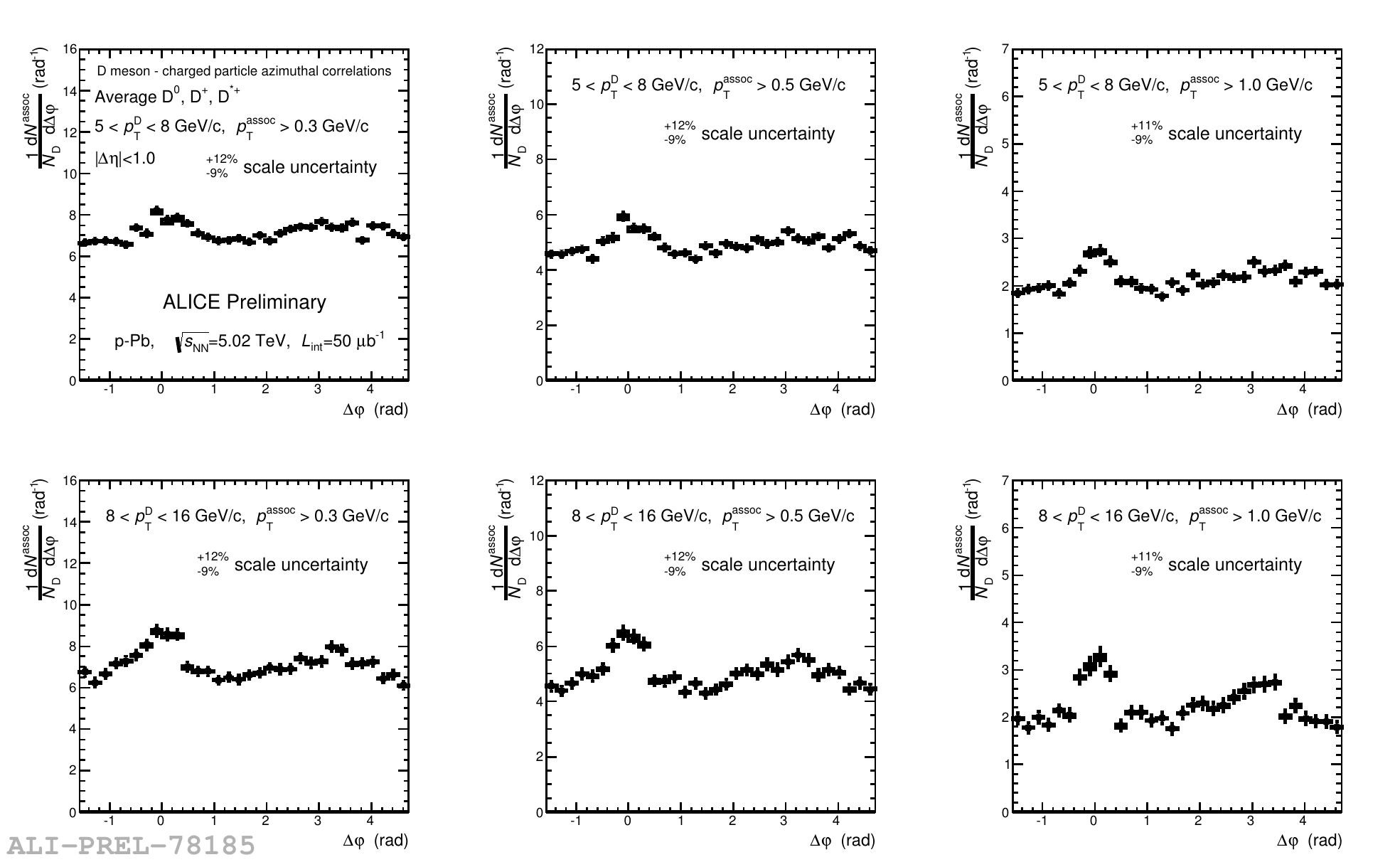}
 \caption{Average azimuthal correlations of $\Dzero$, $\Dplus$ and $\Dstar$ mesons with charged particles in p--Pb collisions at $\sqrtsNN=5.02~\tev$.}
\label{fig:dPhi_Avg_pPb}
\end{figure}
Figure~\ref{fig:dPhi_Avg_pPb} shows the average azimuthal correlation distributions of $\Dzero$, $\Dplus$, $\Dstar$ mesons and charged particles in the interval $5<\ptD<8~\gevc$ (top row) and $8<\ptD<16~\gevc$ (bottom row) for D mesons and with $\ptAssoc>0.3~\gevc$ (letf column), $\ptAssoc>0.5~\gevc$ (middle column), and $\ptAssoc>1~\gevc$ (right column) for the associated charged particles in p--Pb collisions at $\sqrtsNN=5.02~\tev$ with all sorts of corrections mentioned in the section 2. A near-side peak around $\Delta\varphi = 0$ and a wider away-side peak around $\Delta\varphi = \pi$ are clearly visible. A rising trend of near-side peak with the increasing D-meson $\pt$ can be observed.
\begin{figure}[h!]
    \centering
    \begin{minipage}{.5\textwidth}
        \centering
        \includegraphics[width=7cm, height=0.26\textheight]{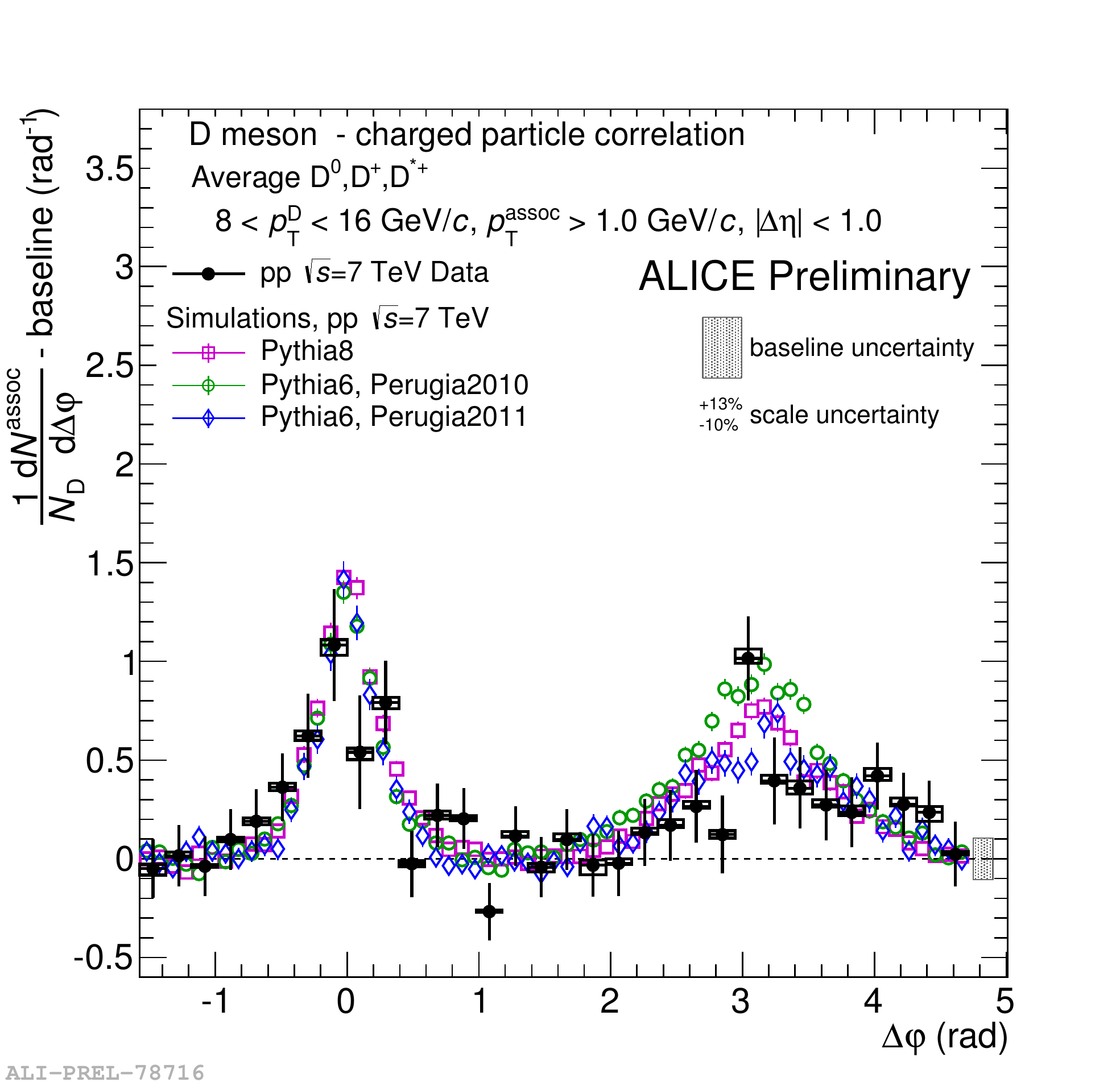}
    \end{minipage}%
    \begin{minipage}{0.5\textwidth}
        \centering
        \includegraphics[width=6.8cm, height=0.26\textheight]{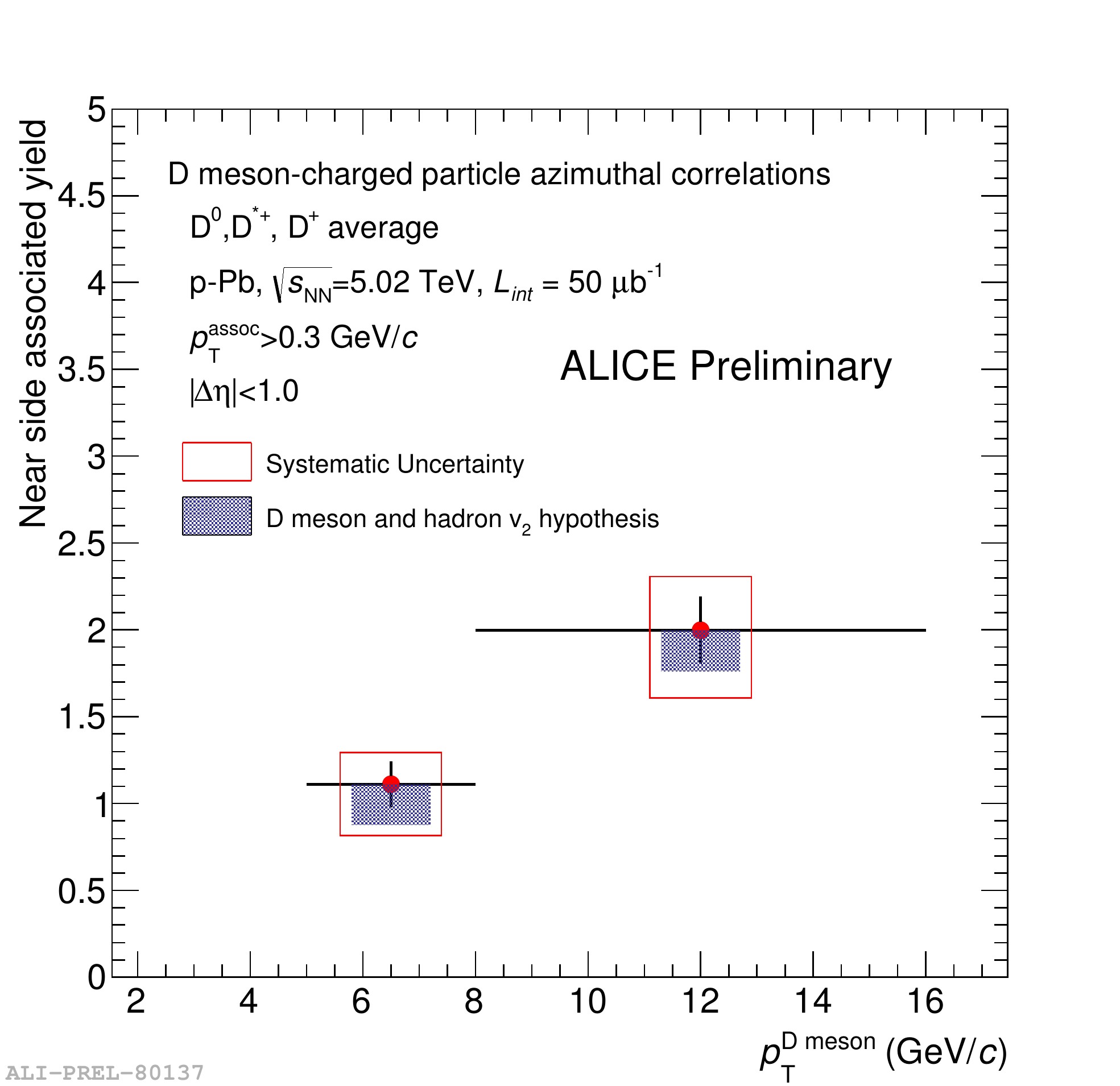}
    \end{minipage}
    \caption{Left: Comparison of average correlation distributions of D mesons and charged particles measured in pp collisions at $\sqrts=7~\tev$ and Monte Carlo simulations performed with PYTHIA, after the subtraction of the baseline. \iffalse The statistical and systematic uncertainties of the measured distributions in the same way\fi Right: Average near-side associated yields as measured from the fit of azimuthal correlation as a function of average D-meson $\pt$ in p--Pb collisions at $\sqrtsNN=5.02~\tev$.}
 \label{fig:NS_Yiled_pPb}		
\end{figure}

Figure~\ref{fig:NS_Yiled_pPb} shows the comparison of the average azimuthal correlation distributions after the subtraction of the baseline measured in pp collisions, with expectations from simulations performed with PYTHIA~\cite{pythia}. The distributions obtained with the different PYTHIA tunes~\cite{PerugiaTunes} considered do not show significant differences in the near side. In the away side, Perugia 2011 tends to have lower correlation values, for $\ptAssoc>1~\gevc$. The different PYTHIA tunes considered in the analysis can describe the data within the measurement uncertainties. 
The azimuthal correlations are fitted with a double Gaussian function and a pedestal, and the near-side associated yields are obtained by integrating the Gaussian distribution, centered at $\Delta\varphi$ = 0. In case of p--Pb a possible $\vtwo$-like modulation (assuming $\vtwo=0.1$ for both D mesons and charged particles) of the baseline is considered for the yield calculations. The right panel of Fig.~\ref{fig:NS_Yiled_pPb} shows the average near-side associated yield for D mesons in $5<\ptD<8~\gevc$ and $8<\ptD<16~\gevc$ with $\ptAssoc>0.3~\gevc$ in p--Pb collisions. They grey box shows the D-meson and charge particle $\vtwo$ hypothesis for the yield calculations. The associated near-side yield increases with  the increase of D-meson $\pt$.
\begin{figure}[h!]
    \centering
    \begin{minipage}{.5\textwidth}
        \centering
        \includegraphics[width=7cm, height=0.26\textheight]{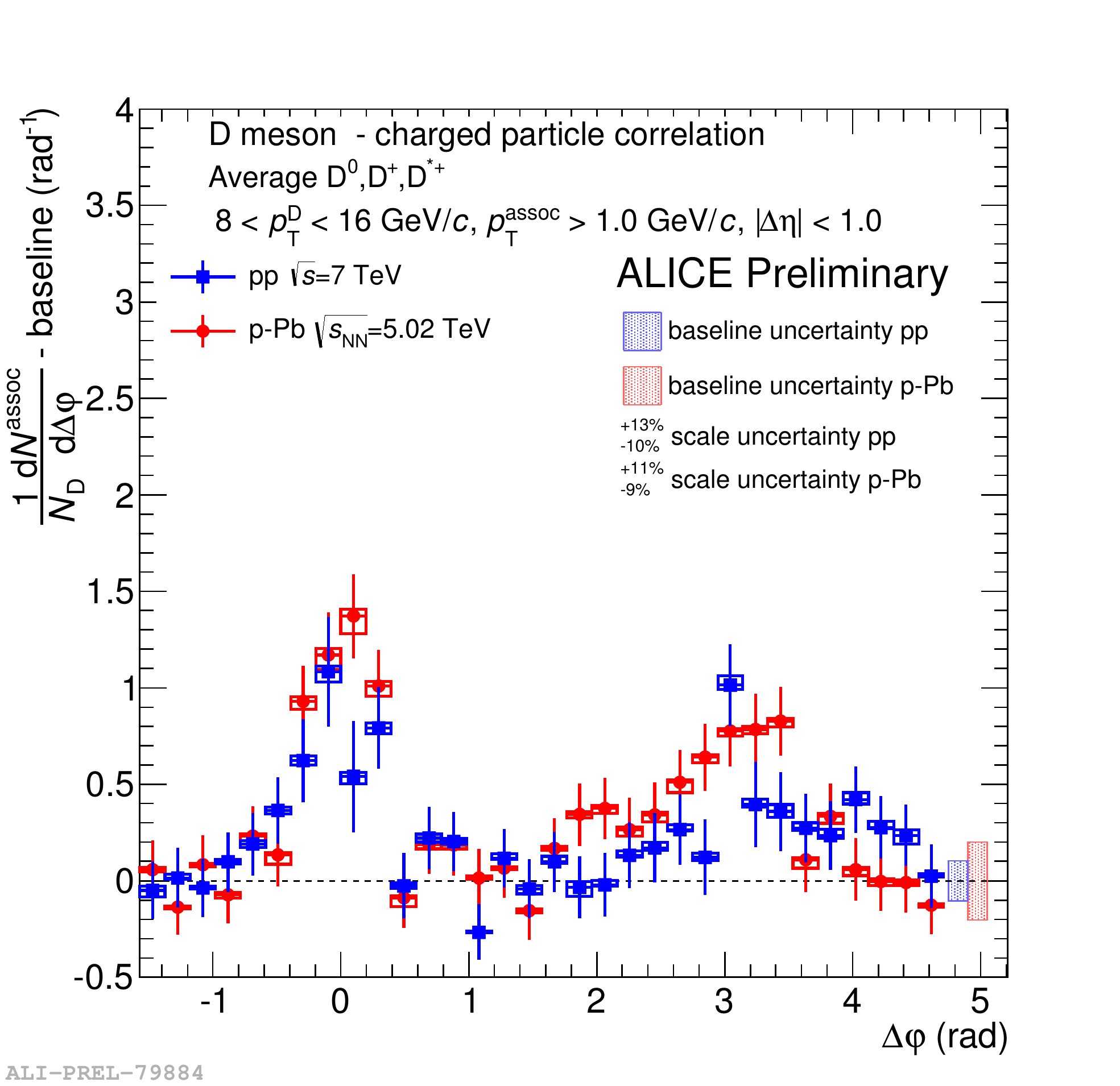}
    \end{minipage}%
    \begin{minipage}{0.5\textwidth}
        \centering
        \includegraphics[width=7cm, height=0.26\textheight]{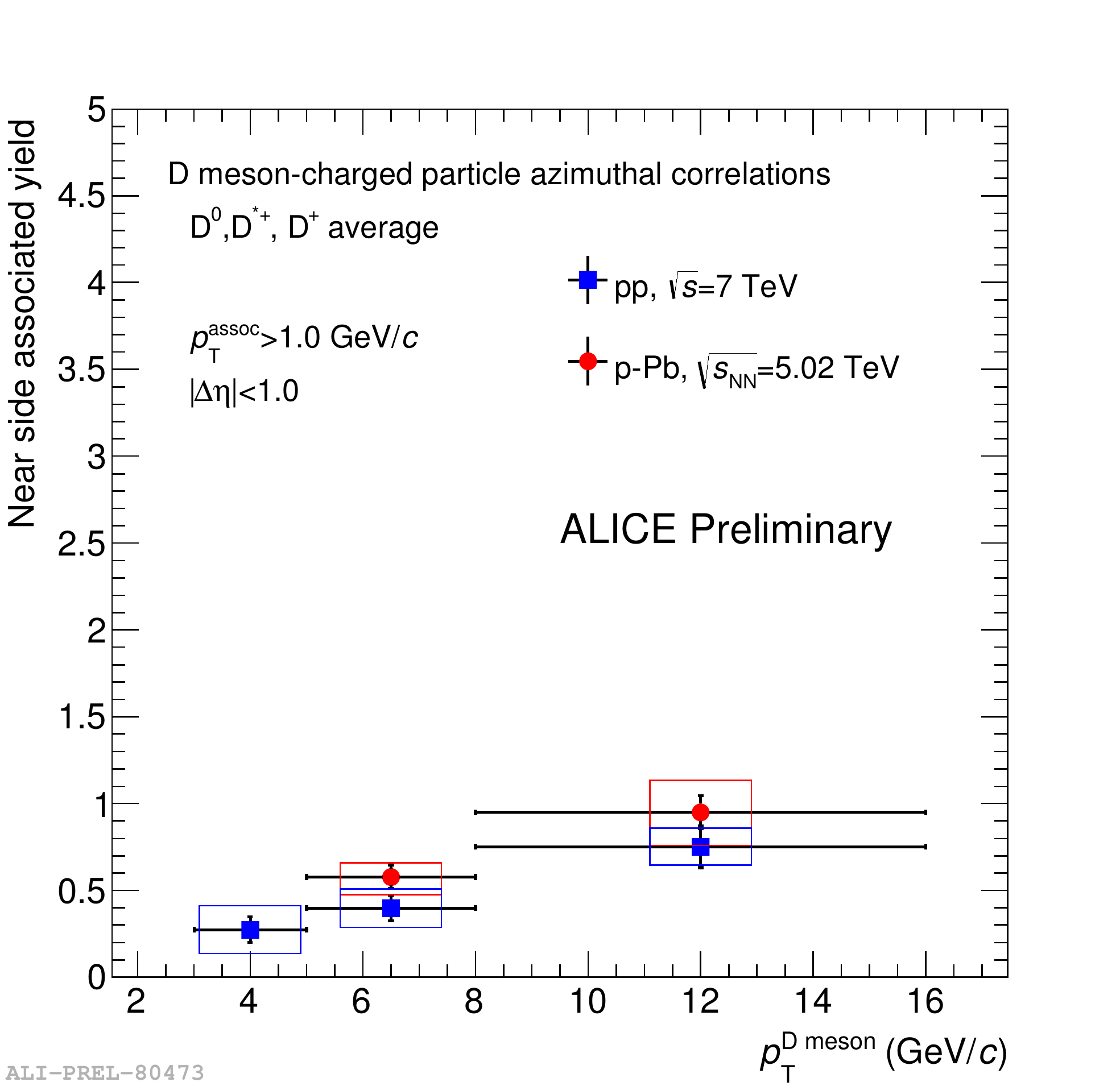}
    \end{minipage}
    	\label{fig:pp_pPbComparison}
    \caption{Left: Comparison of D meson and charged particle correlations in pp at $\sqrts=7~\tev$ and p--Pb at $\sqrtsNN=5.02~\tev$ after baseline subtraction. Right: Comparison of near-side associated yield as function of the D-meson $\pt$ for two collision systems}
\end{figure}
Figure~\ref{fig:pp_pPbComparison} shows the comparison of pp and p--Pb results. The azimuthal correlations for both systems after baseline subtractions are well in agreement within uncertainties. The right panel of Fig.~\ref{fig:pp_pPbComparison} shows the trends of the near-side associated yield as function of the D-meson $\pt$ in pp and p--Pb collisions, confirming the agreement within uncertainties.

%% file: conclusions.tex
We have discussed the results from the first measurement of azimuthal correlations between D mesons and charged particles in pp collisions at $\sqrts=7~\tev$ and p--Pb collisions at $\sqrtsNN=5.02~\tev$, performed with the ALICE experiment at the LHC. The correlations were studied in three different $\pt$ intervals of D mesons and with three different $\pt$ thresholds for the associated primary charged particles. We have seen that within uncertainties the correlation results are compatible for the two collision systems, as expected from simulation studies of pp collisions performed at the two center-of-mass energies. With the current amount of data it is not possible to conclude on any possible modification due to cold nuclear matter effects in p--Pb collisions.\\
In the future, with higher statistics and precision, we can study heavy-flavour azimuthal correlations for heavy-ion data recorded by ALICE at the LHC, which could give further insights into the possible charm energy loss mechanism inside the QGP.